\documentclass[a4paper]{jpconf}

\usepackage{graphicx}
\usepackage{citesort}
\usepackage{revsymb}	
\usepackage{amsmath}	
\usepackage[usenames]{color}	
\usepackage{epstopdf}	
\newcommand{\eq}[1] {Eq.\,(\ref{#1})}

\begin{document}
\title{Amplitudes of solar p modes: \emph{modelling of the eddy time-correlation function}}

\author{K. Belkacem$^{1,2}$, R. Samadi$^{2}$, M.J. Goupil$^{2}$}

\address{$^1$ Institut d'Astrophysique et de G\'eophysique, Universit\'e de Li\`ege, All\'ee du 6 Ao\^ut 17-B 4000 Li\`ege, Belgium.\\
$^2$ LESIA, UMR8109, Universit\'e Pierre et Marie Curie, Universit\'e Denis Diderot, Obs. de Paris, 92195 Meudon Cedex, France.}

\ead{Kevin.Belkacem@ulg.ac.be}

\begin{abstract}
Modelling amplitudes of stochastically excited oscillations in stars is a powerful tool for understanding the properties of the convective zones. For instance, it gives us information on the way turbulent eddies are temporally correlated in a very large Reynolds number regime. 
We discuss the way the time correlation between eddies is modelled and we present recent theoretical developments as well as  observational results. Eventually, we discuss the physical underlying meaning of the results by introducing the Ornstein-Uhlenbeck process, which is a sub-class of a Gaussian Markov process. 
\end{abstract}

\section{Introduction}
\label{intro}
\vspace{0.3cm}

Amplitudes of solar-like oscillations  result from a balance between excitation and damping 
and crucially depend on the way the eddies are temporally correlated (see the review of \cite{Samadi09}). 
Hence, the improvement of our understanding and modeling of the temporal correlation of turbulent eddies, hereafter denoted in the Fourier domain as $\chi_k(\omega)$, is fundamental to infer turbulent properties in stellar convection zones. 

Most of the theoretical formulations of mode excitation explicitly or implicitly assume a Gaussian functional form for $\chi_k(\omega)$ (\cite{GK77,Dolginov84,GK94,B92c,Samadi00II,Chaplin05}). 
However, 3D hydrodynamical simulations of the outer layers
of the Sun show that, at the length-scales close to that of the energy
bearing eddies (around $1$ Mm), $\chi_k$ is a Lorentzian function  \cite{Samadi02II,Belkacem09}.  As pointed-out by \cite{Chaplin05}, a Lorentzian  $\chi_k$  is also a result predicted for the largest, most-energetic eddies by the
time-dependent mixing-length formulation derived by \cite{Gough77}.  However, \cite{Chaplin05}, \cite{Samadi09}, and \cite{Houdek09} found that a Lorentzian $\chi_k$, when used with a mixing-length description of the whole convection zone, results in a severe over-estimation for the low-frequency modes. 
 
In a recent work, \cite{Belkacem11} introduced a cut-off frequency in the modelling of the eddy-time correlation function to account for the effect of short-time scales. Indeed, under the sweeping approximation, which consists in assuming that the temporal correlation of the eddies, (in the inertial subrange) is dominated by the advection by energy-bearing eddies, the shape of the temporal correlation function of eddies is no longer Lorentzian. 

Hence, in this paper we discuss the modelling of the eddy-time correlation function in both the limit of large and short time scales. This modelling is then validated using observational constraints from amplitudes of solar-like oscillations. Eventually, the underlying physical meaning of the observed shape of the temporal correlation function is discussed by mean of a simplified stochastic process. 

\section{Modelling the  Eulerian eddy time-correlation function} 
\label{chiw}

\vspace{0.3cm}

The formalism we used to compute excitation rates of radial modes was developed by \cite{Samadi00I} (see also \cite{Samadi05c}). For a radial mode of frequency $\omega_0=2\pi \, \nu_0$, the excitation rate (or equivalently, the energy injection rate), $P$,  mostly  arises from the Reynolds stresses  and can be written  as (see Eq.~(21) of \cite{Belkacem08})  
\begin{align}
\label{puissance}
P(\omega_0) &= \frac{\pi^{3}}{2  I} \int_{0}^{M} \, \left[ \rho_0  \, \left(\frac{16}{15}\right) \left(\frac{\partial \xi_r}{\partial r}\right)^2 \;  \int_{0}^{+\infty}  
 \mathcal{S}_k \; \textrm{d}k \right]  \textrm{d}m \\
\label{source}
\mathcal{S}_k &= \frac{E^2(k)}{k^2 } \int_{-\infty}^{+\infty}   
~\chi_k( \omega + \omega_0) ~\chi_k( \omega ) \; \textrm{d}\omega
\end{align}
where $\xi_r$ is the radial component of the fluid displacement eigenfunction ($\vec \xi$), $m$ is the local mass, $\rho_0$ the mean density, $\omega_0$ the mode angular 
frequency, $I$ the mode inertia,
$\mathcal{S}_k$ the source function,  $E(k)$ the spatial kinetic energy 
spectrum, $\chi_k$ the eddy-time correlation function, and $k$ the wave-number. 

\subsection{Some preliminary definitions}
\vspace{0.3cm}

We now turn to a rigourous definition of the eddy-time correlation function ($\chi_k$). 
For a turbulent fluid, one defines the Eulerian eddy time-correlation function as
\begin{align}
 \left< \vec u(\vec x + \vec r,t+\tau) \cdot \vec u(\vec x,t) \right> = \int  {\cal E}(\vec k,t,\tau) \, e^{{\rm i} \vec k \cdot \vec x} \, {\rm d}^3\vec k  \, ,
\label{time-correlation}
\end{align}
where $\vec u$ is the Eulerian turbulent velocity field, $\vec x$ and $t$ the space and time position of the fluid element, $\vec k$ the wave number vector, $\tau$ the time-correlation length, and $\vec r$ the space-correlation length. 
The function $\cal E$ in the RHS of \eq{time-correlation} represents the  time-correlation function associated with an eddy of wave-number $\vec k$.

We assume an isotropic and stationary turbulence, accordingly  ${\cal E}$ is only a function of $k$ and $\tau$.
The quantity ${\cal E}(k,\tau)$ is related to the turbulent energy spectrum according to
\begin{align}
{\cal E}(k,\tau) = \frac{E(k,\tau)}{2\pi k^2} \, .
\end{align}
where $E(k,\tau)$ is the turbulent kinetic energy spectrum whose temporal Fourier transform is 
\begin{align}
\label{Ekw}
E(k,\omega)   \equiv {1 \over {2\pi}} \,  \int_{-\infty}^{+\infty}  {E}(k,\tau) \, e^{{\rm i} \omega \tau} \, {\rm d} \tau
\end{align}
where $\omega$ is the eddy frequency, and $E(k,\omega)$ is written as follows \cite{Stein67,Samadi00I}  
\begin{align}
\label{decomp_E}
E(k,\omega) = E(k) \, \chi_k (\omega) \quad {\rm with}\quad  \int_{-\infty}^{+\infty} \chi_k (\omega) \, {\rm d} \omega = 1
\end{align}
where $\chi_k(\omega)$ is the frequency component of $E(k,\omega)$. In other
words, $\chi_k(\omega) $  represents -~  in the frequency  domain ~-  the temporal correlation between eddies of wave-number $k$. 

\subsection{Long-time scales}
\vspace{0.3cm}

A modelling of $\chi_k$ in the stellar context from the full non-linear hydrodynamic equations is a very difficult task, and no such a model has been proposed so far. This is particularly the case for large times, which we defines by the times of the order or greater than the integral time-scale that is the width at half maximum of $\chi_k$. Hence,  we adopt an alternative approach that is to postulate the functional form of $\chi_k$ and then to confront it to observations and/or 3D hydrodynamical numerical simulations.  

Several functional have then been tested in the past and as mentioned in Sect.~\ref{intro}, theoretical and observational evidence show that $\chi_k(\omega)$ is Lorentzian, \emph{i.e.} 
\begin{align}
\chi_k (\omega)  = { 1 \over {\pi \omega_k} } \, {1 \over {1 + \left ( \omega/\omega_k \right )^2} } 
\quad {\rm with} \quad \omega_k = k \, u_k \quad {\rm and}\quad u_k^2 = \int_{k}^{2k} E(k) \, {\rm d}k
\label{lorentzian}
\end{align}
where $\omega_k$ is by definition the width at half maximum of $\chi_k(\omega)$. In the framework of \cite{Samadi00I}'s formalism, this latter quantity is evaluated as:
\begin{align}
\label{omega_k}
\omega_k = k \, u_k \quad {\rm with}\quad u_k^2 = \int_{k}^{2k} E(k) \, {\rm d}k
\end{align} 
where $E(k)$ is defined by \eq{decomp_E}. Note that \eq{lorentzian} corresponds to an exponential in the time domain. 

Such a modelling of the temporal correlation of eddies has successfully been tested in the solar and stellar context (see \cite{Samadi09} for a detailed discussion). However, when coupled with a standard 1D model of the solar convection zone, it leads to an important overestimation of the solar mode amplitudes. More precisely, a substantial fraction of the energy injected to those  modes comes from the deep layers of the solar convective region. As a result, \cite{Chaplin05} and \cite{Samadi09} then suggested that most contributing eddies situated deep in the Sun have a $\chi_k$ rather Gaussian than Lorentzian since, at fixed frequency, a Gaussian $\chi_k$ decreases more rapidly with depth. Nevertheless, this conclusion has recently been questioning by \cite{Belkacem11}, which demonstrate that \eq{lorentzian} is no longer valid for short-time scales. 

\subsection{Short-time scales}
\label{short_time}
\vspace{0.3cm}

To infer some properties of the eddy-time correlation function at short-time scales, the function ${\cal E}(k,t,\tau)$ appearing in \eq{time-correlation} is expanded for short time scales. In the inertial sub-range it gives (see \cite{Kaneda93} for a detailed derivation)
\begin{align}
\label{DL}
{\cal E}(k,\tau) = {\cal E} (k,\tau=0) \, \left ( 1  -  \alpha_k  {\left | \tau  \right |}  -  \frac{1}{2} \,  \left ( { \omega_E \tau } \right )^2  + \dots \right )
\end{align}
where the  characteristic frequency $\alpha_k$ can be estimated by the eddy turn-over frequency $\omega_k$ (see \cite{Belkacem11} for details), and 
the second characteristic frequency, $\omega_E (k)$, is the curvature of the correlation function near the origin, and defined by 
\begin{align}
\label{tau_E}
\omega_E = k \, u_0 \, .
\end{align}
This expression is obtained under the sweeping assumption, which consists in assuming that the velocity field $\vec{u}(k)$ associated with an eddy of  wave-number $\vec k$ lying in the inertial-subrange (\emph{i.e.} large $\vec k$ compared to $\vec k_0$) is advected by  the energy-bearing eddies with velocity $\vec{u}_0$ (\emph{i.e.} of wave-number $\vec k_0$). 

To go further, we note that the zeroth- and first-order terms in \eq{DL} are consistent with an exponential decrease of width $\alpha_k$ (\emph{i.e.} a Lorentzian in the frequency domain of width $\omega_k$, \eq{lorentzian}) for small $\tau$. In contrast, the zeroth-order term together with the second order term in \eq{DL} are consistent with a Gaussian behavior of width $\tau_E$.   
In turn, \cite{Belkacem11} have shown that for $\omega > \omega_E$ the second order term dominates over the first order one in \eq{DL}, at all  length-scales.  Hence, for frequencies near the micro-scale frequency ($\omega > \omega_E$), the eddy-time correlation function behaves as a Gaussian function $(e^{-(\omega/\omega_E)^2})$ rather than a Lorentzian function, resulting in a sharp decrease with $\omega$. Hence, the contributions for $\omega > \omega_E$ are negligible. 

\section{Observational evidences for a Lorentzian eddy-time correlation function}
\vspace{0.3cm}

\begin{figure}[t]
\begin{center}
\includegraphics[height=5.5cm,width=7.5cm]{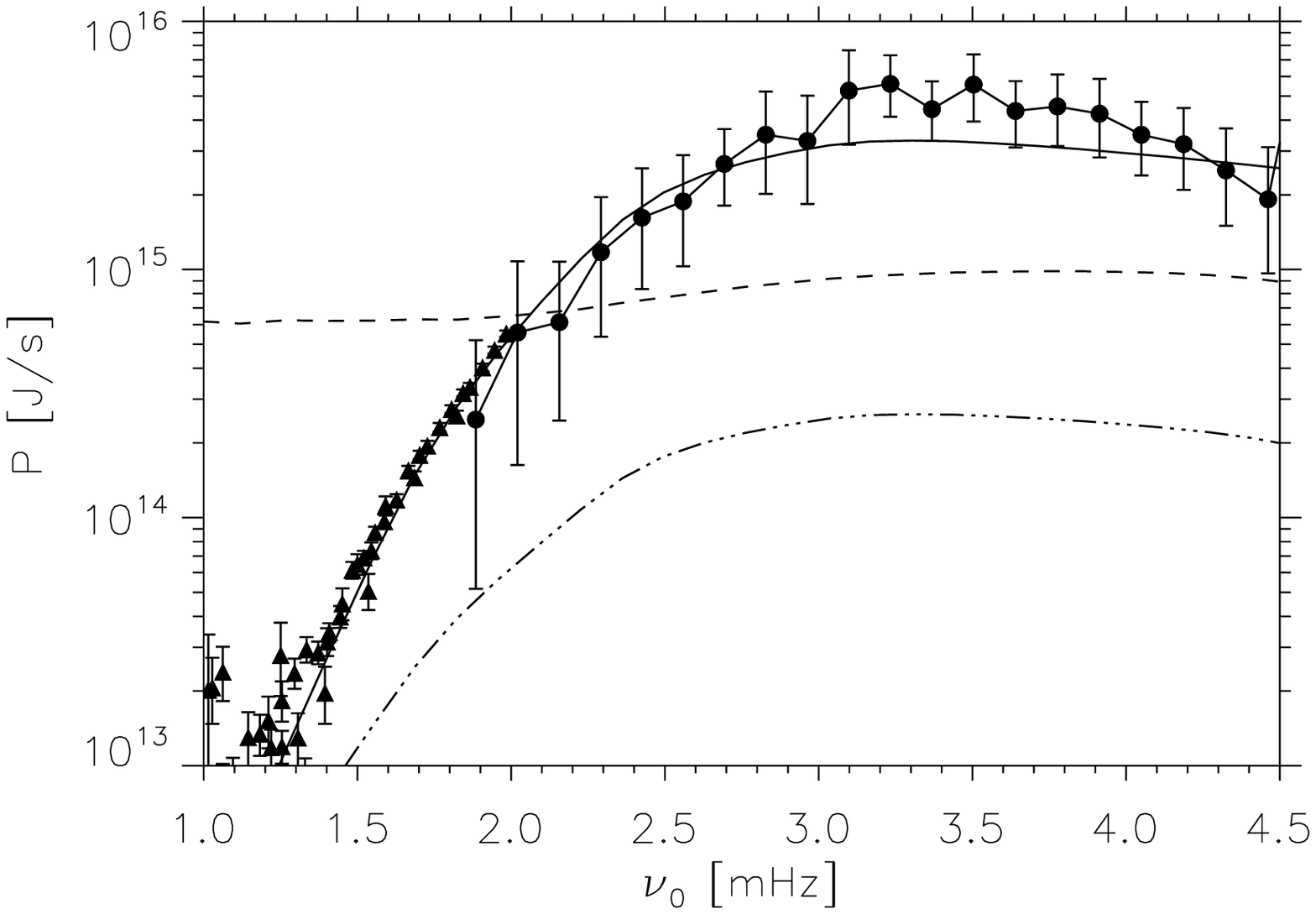}
\includegraphics[height=5.5cm,width=7.5cm]{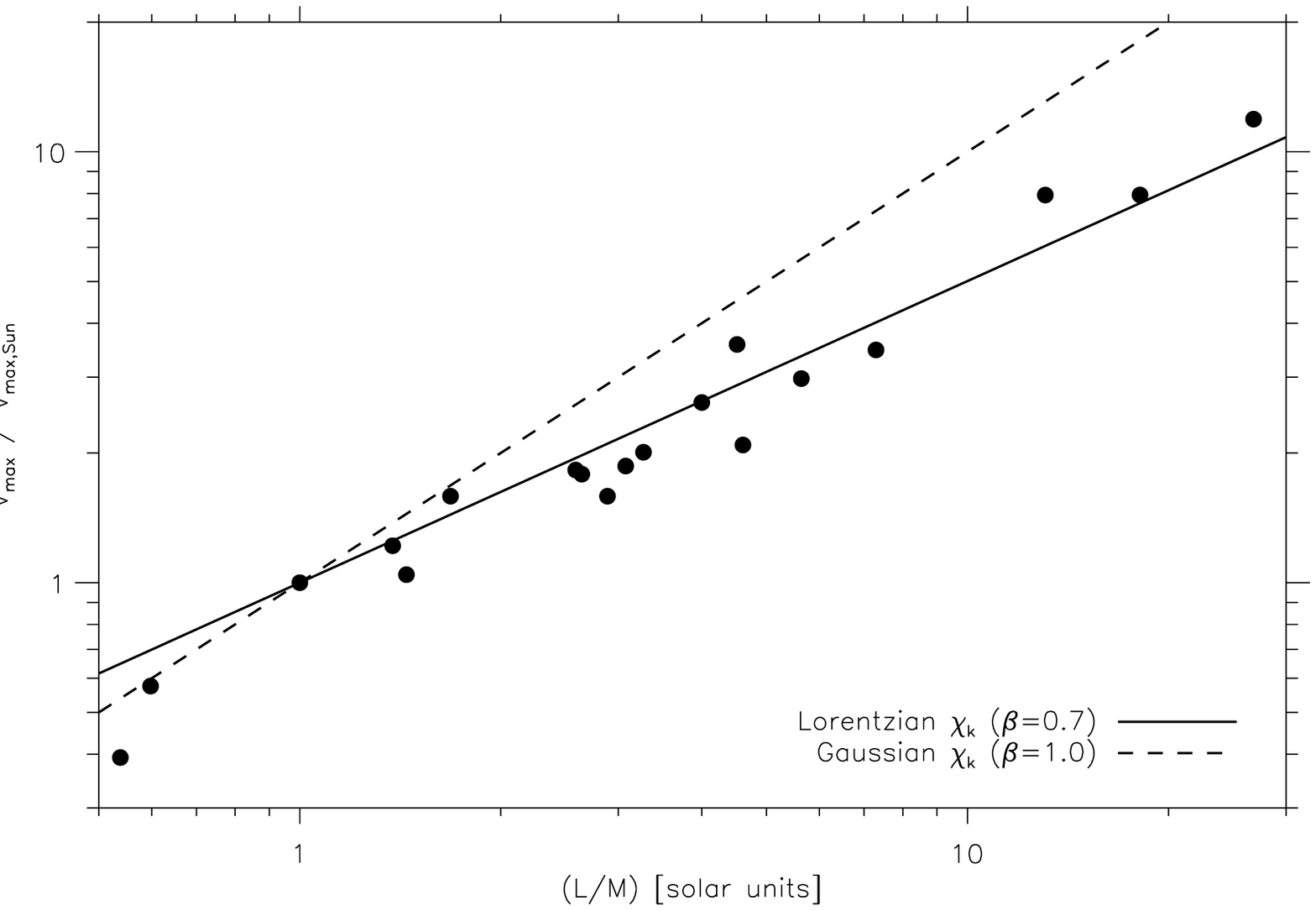}
\caption{{\bf Left panel:} 
Solar $p$-mode excitation rates as a function of the frequency $\nu$. The dots correspond to the observational data obtained by the GONG network, as derived by \cite{Baudin05}, and the triangles corresponds to observational data obtained by the GONG network as derived by \cite{Salabert09} for $\ell=0$ to $\ell=35$. 
The dashed line corresponds to the computation of the excitation rates using with a Lorentzien description of $\chi_k$ \emph{without} any cut-off frequency, while the dashed triple dot line corresponds to the same computation except that the Lorentzian description of $\chi_k$ down to the cut-off frequency $\omega_E$.  
The solid line corresponds to the computation of mode excitation rates using 3D numerical computation to described turbulent convection properties and a Lorentzian $\chi_k$ together \emph{with} a cut-off frequency at $\omega=\omega_E$. Note that no scaling is applied to those calculations. 
{\bf Right panel:} Ratio between ($V_{\rm max}$) the maximum of the mode velocity relative to the observed solar value ($V_{\odot,max} = 25.2 cm/s$ for $\ell$ = 1 modes, see Kjeldsen et al. 2008). Filled dots correspond to the stars for which solar-like oscillations have been detected in Doppler velocity (see \cite{Bedding07}. The lines  correspond to the power laws obtained from the predicted scaling laws for $P_{max}$ and estimated values of the damping rates $\eta_{max}$ (see \cite{Samadi09} for details).}
\label{resultat}
\end{center}
\end{figure}

\subsection{The solar $p$-mode energy injection rates}
\label{amplitude}
\vspace{0.3cm}

When the frequency range of $\chi_k$ is extended toward infinity, computation of $P$ fails to reproduce the observations, in particular the low-frequency shape (see Fig.~\ref{resultat}). It is in agreement with the results of \cite{Chaplin05} and \cite{Samadi09}, and results in an over-estimation of the excitation rates at low frequency.  
In contrast, by assuming that the time-dynamic of eddies in the Eulerian point of view is dominated by the sweeping, the Eulerian time micro-scale arises as a cut-off frequency (see Sect.~\ref{short_time}). Hence, $\chi_k (\omega)$ is modeled following \eq{lorentzian} for $\omega < \omega_E$ and $\chi_k (\omega) = 0$ elsewhere. 
Using such a procedure to model $\chi_k(\omega)$ permits us to reproduce the low-frequency ($\nu < 3$mHz) shape of the mode excitation rates as observed by the GONG network (see Fig.~\ref{resultat}).  
This is explained as follows; for large scale eddies near $k_0^{-1}$, situated deep in the convective region, the cut-off frequency $\omega_E$  is close to $\omega_k$. As a consequence, the frequency range over which $\chi_k$ is integrated in \eq{source} is limited, resulting in lower injection rates into the modes. 
This results then re-conciliates theoretical and observational evidences that the frequency dependence of the eddy-time correlation must be Lorentzian in the whole solar convective region down to the cut-off frequency $\omega_E$.

 Note that a modelling of the excitation rates, using a Gaussian for $\chi_k$ and a 3D numerical simulation to model convection, leads to an underestimation of the excitation rates, as shown by \cite{Belkacem11}. 

\subsection{Scaling law on the maximum mode energy injection rate for solar-type pulsators}
\label{scaling_vmax}
\vspace{0.3cm}

A similar conclusion has been reached by \cite{Samadi07} for main-sequence and 
red giant stars. Indeed, \cite{Houdek1999} have shown that the maximum mode surface velocity scales as a function of the ratio of luminosity to the mass ($L/M$). Later, \cite{Samadi07} have shown that the exponent $\beta$ 
is sensitve to the adopted model for computing $\chi_k$, \emph{i.e.} $\beta=0.7$ for a Lorentzian and $\beta=1$ for a Gaussian.  As shown in Fig.~\ref{resultat}, the best agreement with the observations is found when a 
Lorentzian $\chi_k$ is assumed. In contrast, assuming a Gaussian $\chi_k$ results 
in a larger exponent $\beta$ and do not permit to reproduce the observations. 
Consequently, we conclude that the eddy-time correlation function is a Lorenztian functional (or equivalently an exponential functional in the time domain) for all solar-like pulsators and in the upper-layers of convective regions.  Note that the case of deep layers in convective regions is still to be investigated, but 3D numerical results from \cite{Belkacem09} also suggest $\chi_k$ is Lorentzian.

\section{On the interpretation of the Lorentzian shape of $\chi_k$}
\vspace{0.3cm}

A modelling of the Eulerian time correlation function from the full non-linear hydrodynamic equations is a difficult task, making the interpretation of our results non-trivial. However, one can use a classical approach in turbulence, which consists in modelling the turbulence by a stochastic model (see the review \cite{Pope94}). Similarly to the Brownian motion, one can define a fluid elements and express the corresponding Langevin equation. Naturally, a Lagrangian approach is more adapted however, our modelling of mode amplitudes deals with Eulerian velocities. To overcome this difficulty we first introduced a relation between Eulerian and Lagrangian time correlation function, first introduced by \cite{Corrsin59}. 

\subsection{The Corrsin's conjecture}
\vspace{0.3cm}

The problem of relating Eulerian and Lagrangian correlation is an outstanding problem in turbulence, since the measurement of Lagrangian quantities is experimentally difficult. A general expression can be written such as 
\begin{align}
\label{Corrsin01}
\left < w(t+\tau) w(t) \right> =  \int {\rm d}^3\vec r \left< v(\vec x+\vec r,t+\tau) v(\vec x,t) \psi(\vec x,t) \right> 
\end{align}
where $w$ is the Lagrangian velocity and $u$ the Eulerian velocity, and $\psi$ is the probability distribution function of the fluid particule displacement at time $t$. 
In this framework, \cite{Corrsin59} proposed a simplified relation based on the assumption that for large times ($\tau$), $\psi$ and $u$ are statistically  weakly dependent. Then by using a Gaussian distribution for $\psi$, \cite{Corrsin59} demonstrates that one can rewritte \eq{Corrsin01} in the limit of large scales\footnote{This assumption is valid for our purposes  since modes are mainly excited by large scales. }
\begin{align}
\label{Corrsin03}
\left < w(t+\tau) w(t) \right> \approx \left< v(\vec x,t+\tau) v(\vec x,t) \right> 
\end{align}
This approximated relation has been extensively discussed and has been shown to be rather accurate for turbulent flows without helicity (e.g. \cite{Kraichnan77,Ott05}).  

\subsection{The Ornstein-Uhlenbeck Process}
\vspace{0.3cm}

For long time-scales, given the relation \eq{Corrsin03}, we now turn to discuss some aspect of Lagrangian stochastic models. In the limit of a stationary, homogeneous turbulence, 
a stochastic model can be adopted that follows the Langevin equation 
(written in term of stochastic differential equation)
\begin{align}
\label{stoch}
{\rm d}w(t) = - w(t) {\rm d}t / T + (2u^{\prime 2}/T) \zeta(t) {\rm d}t 
\end{align}
where $T$ is the Lagrangian integral time-scale, $u^{\prime 2}$ is the variance of $w(t)$, and $\zeta$ a random fluctuating term. To go further, lets us assume that $\zeta$ is a Gaussian random variable with zero mean and which is rapidly varying, \emph{i.e.} $<\zeta(t) \zeta(t^\prime)>=\delta (t-t^\prime)$. 

Consequently, the Langevin equation \eq{stoch} describes a continuous Gaussian Markov process in time (it is a Wiener process or equivalently a Brownian process). More precisely, since we are interested to describe a stationary process, $w(t)$ is a stationary, Gaussian, Markov process know as the Ornstein-Uhlenbeck Process (see \cite{Gardiner90} for a more detailed definition). In other terms it corresponds to a noisy relaxation process. A general property of such a process is that its time correlation function is (e.g., \cite{Gardiner90,Pope94})
\begin{align}
\label{corr1}
\frac{\left < w(t+\tau) w(t) \right>}{u^{\prime 2}} = e^{-| \tau | / T}
\end{align}
which, in the time Fourier space, corresponds to a Lorentzian function. 

A more rigourous investigation has been proposed by \cite{Sawford91}, who proposed a stochastic model for turbulent dissipation. The author then proposed an expression for the time correlation function as a function of the Reynolds number
\begin{align}
\frac{\left < w(t+\tau) w(t) \right>}{u^{\prime 2}} =  \left[ e^{-| \tau | / T} - Re^{-1/2} e^{- Re^{-1/2} | \tau | / T} (1-Re^{-1/2})^{-1}  \right]
\end{align}
which tends to \eq{corr1} for $Re \rightarrow +\infty$.

Hence, by used of the Corrsin's conjecture one can conclude that our results, obtained using mode amplitudes (\emph{i.e.} $\chi_k$ is Lorentzian in the time Fourier domain), suggests that turbulent eddies in convective region of solar-type stars follow such a noisy relaxation process. 

\section{Conclusion}
\label{conclu}
\vspace{0.3cm}

We have shown that the modelling of mode amplitudes gives us access to some properties of turbulent convection in stars. More precisely, we detailed the way the time correlation between eddies is actually modelled and validated 
using observational data. The conclusion is that the eddy-time correlation function follows a Lorentzian shape (in the time Fourier domain) down to a cut-off frequency, in convective region of solar-type stars. 

We then discussed this shape and we emphasised that a Lorentzian function - or equivalently an exponential function in the time domain - is a rather common result for noisy relaxation processes.  Our discussion also shows that maybe the more interesting will be to investigate the departure from a Lorentzian shape of $\chi_k$, and its physical interpretation. To this end, we stress that more accurate observations for the Sun of solar-like pulsators is mandatory. 

\section{References}

\bibliographystyle{iopart-num}
\bibliography{bib.bib}

\end{document}